\documentclass[aps,pra,10pt,twocolumn,superscriptaddress,floatfix]{revtex4-2}
\usepackage[bookmarks=true,colorlinks,linkcolor=RoyalBlue,urlcolor=NavyBlue,citecolor=RoyalBlue]{hyperref}
\usepackage{amsmath,amssymb}
\usepackage{upgreek}
\usepackage{braket}
\usepackage{mathtools}
\usepackage[english]{babel}
\usepackage{graphicx}
\usepackage[dvipsnames]{xcolor}

\newcommand{\euler}{\mathrm{e}}
\newcommand{\imunit}{\mathrm{i}}
\newcommand{\boldone}{\text{\usefont{U}{bbold}{m}{n}1}}

\begin{document}

\title{Probing confinement in a \texorpdfstring{$\mathbb{Z}_2$}{Z2} lattice gauge theory on a quantum computer}
\author{Julius Mildenberger}
\email{julius.mildenberger@unitn.it}
\affiliation{Pitaevskii BEC Center, CNR-INO and Department of Physics, University of Trento, Italy}
\affiliation{INFN-TIFPA, Trento Institute for Fundamental Physics and Applications, Trento, Italy}
\author{Wojciech Mruczkiewicz}
\affiliation{Google Quantum AI, Venice, California, USA}
\author{Jad C.~Halimeh}
\affiliation{Pitaevskii BEC Center, CNR-INO and Department of Physics, University of Trento, Italy}
\affiliation{Department of Physics and Arnold Sommerfeld Center for Theoretical Physics (ASC), Ludwig-Maximilians-Universit\"at M\"unchen, Germany}
\affiliation{Munich Center for Quantum Science and Technology (MCQST), M\"unchen, Germany}
\author{Zhang Jiang}
\affiliation{Google Quantum AI, Venice, California, USA}
\author{Philipp Hauke}
\email{philipp.hauke@unitn.it} 
\affiliation{Pitaevskii BEC Center, CNR-INO and Department of Physics, University of Trento, Italy}
\affiliation{INFN-TIFPA, Trento Institute for Fundamental Physics and Applications, Trento, Italy}

\vspace*{.01\baselineskip}
\maketitle

\begin{figure*}[t!]
    \centering
    \includegraphics[width=0.99\textwidth]{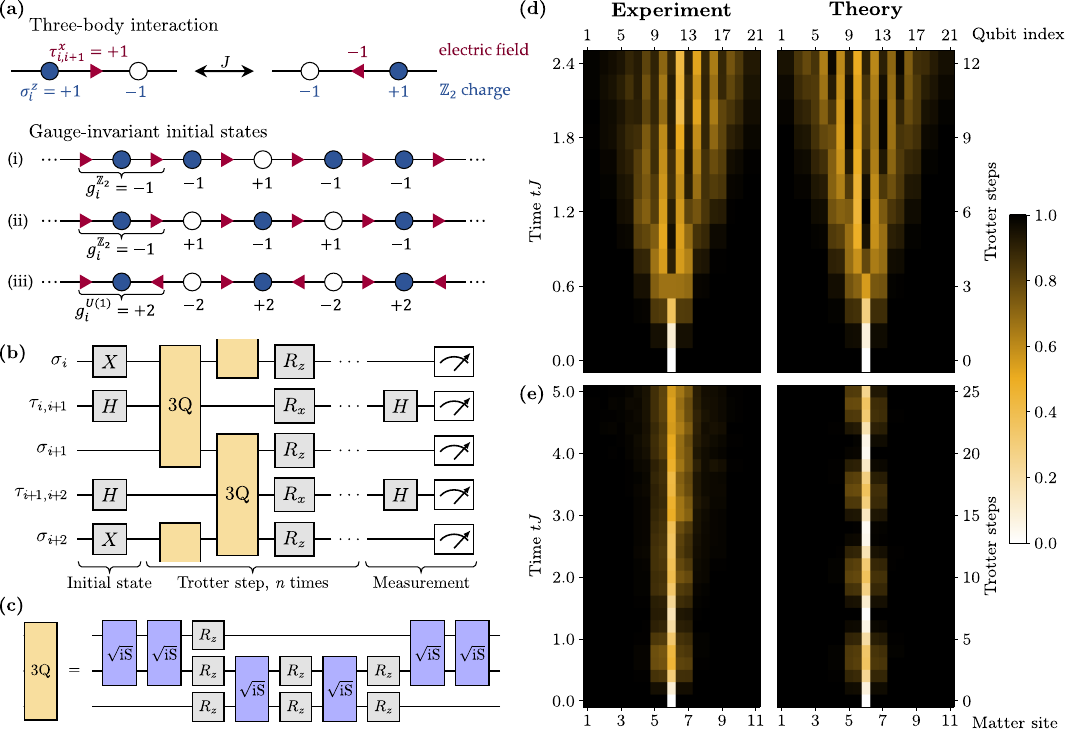}
    \caption{\textbf{Simulating confinement in a $\mathbb{Z}_2$ lattice gauge theory on a digital superconducting chip.} 
    (a) Target gauge theory. Charged matter (Pauli operators $\sigma_i$) lives on sites $i$ and gauge fields (Pauli operators $\tau_{i,i+1}$) on links. 
    The dynamics of charges and gauge field is coupled by a three-qubit (3Q) interaction term $H_J$ and constrained by a $\mathbb{Z}_2$ gauge symmetry, embodied in the conservation of the local generators $G_i^{\mathbb{Z}_2}=-\tau_{i-1,i}^x \sigma_i^z \tau_{i,i+1}^x$. Our simulations use three different types of gauge-invariant initial states. 
    (b) Circuit. The dynamics is implemented by Trotterizing the Hamiltonian into gauge--matter interaction $H_J$, the matter rest mass (realized by $z$-rotations) and a background electric field ($x$-rotations), for a total circuit depth of 8 two-qubit layers per Trotter step. 
    Some of our experiments add a term that protects a $\mathrm{U}(1)$ gauge symmetry by single-qubit rotations, permitting us to tune the system between a $\mathbb{Z}_2$ and $\mathrm{U}(1)$ gauge theory. These can be incorporated into the single-qubit $x$-/$z$-rotations already employed to implement the matter rest mass and background electric field, thereby not increasing gate depth.
    (c) Three-qubit matter--gauge-field interaction. The experiment is enabled by synthesizing the three-qubit interaction in a (in absence of error terms) fully gauge-invariant manner using only 6 two-qubit gates and 3 layers of single-qubit gates. As $z$-rotations, the latter do not require execution time (Methods). 
    (d,e) Confinement dynamics. Plotted is the inverse excitation of qubits at readout, which for odd qubit numbers gives $(\boldone+\sigma_i^z)/2$, corresponding to the $\mathbb{Z}_2$ charge $-\sigma_i^z$, and on even qubit numbers $(\boldone+\tau_{i,i+1}^x)/2$ corresponding to the electric field $\tau_{i,i+1}^x$.
    At small background field strength $f$, a matter defect can spread over the entire finite system (d), while with increasing $f$ it is more strongly confined, purely due to the interaction with the gauge field (e).
    Theory predictions from exact Trotterized numerical simulations agree well with the experimental data from the superconducting Sycamore-class chip.
    }
    \label{fig:Overview}
\end{figure*}

\textbf{
Gauge theories describe the fundamental forces in the standard model of particle physics and play an important role in condensed matter physics. The constituents of gauge theories, for example charged matter and electric gauge field, are governed by local gauge constraints, which lead to key phenomena such as confinement of particles that are not fully understood. In this context, quantum simulators may address questions that are challenging for classical methods. While engineering gauge constraints is highly demanding, recent advances in quantum computing are beginning to enable digital quantum simulations of gauge theories. 
Here, we simulate confinement dynamics in a $\mathbb{Z}_2$ lattice gauge theory on a superconducting quantum processor. Tuning a term that couples only to the electric field produces confinement of charges, a manifestation of the tight bond that the gauge constraint generates between both. Moreover, we show how a modification of the gauge constraint from $\mathbb{Z}_2$ towards $\mathrm{U}(1)$ symmetry freezes the system dynamics. Our work illustrates the restriction that the underlying gauge constraint imposes on the dynamics of a lattice gauge theory, it showcases how gauge constraints can be modified and protected, and it promotes the study of other models governed by multi-body interactions. 
}

The rich phenomenology of gauge theories is fundamentally linked with local constraints set by the underlying gauge symmetry, which determine the interplay between matter and gauge degrees of freedom. 
For example, the non-Abelian gauge symmetry of quantum chromodynamics confines quarks together with gluons into hadrons due to their mutual interaction, and quantum electrodynamics (QED) hosts a local $\mathrm{U}(1)$ symmetry, the well-known Gauss's law, that is responsible for the massless photon and the emergence of the long-ranged Coulomb law \cite{Gattringer2010,Pasechnik2021}. 
Originally employed to address non-perturbative regimes in particle physics \cite{Gattringer2010}, lattice discretizations of gauge theories have proven to be powerful frameworks to describe the emergence of exotic phenomena also in condensed-matter physics including, e.g., disorder-free localization \cite{Disorder-Free_Localization_2}, quantum many-body scars \cite{Aramthottil2022}, and staircase prethermalization \cite{Staircase_Prethermalization_2}.

Among the most fascinating phenomena in lattice gauge theories (LGTs) is confinement, by which the system dynamics is restricted due to interactions that are mediated via the gauge field. 
Confinement can give rise to bound meson excitations and in some cases is even associated with topological phases with non-Abelian anyons and charge fractionalization \cite{Pasechnik2021}. 
In the presence of both dynamical matter and gauge fields, the confinement problem is not fully understood, in part because of the subtle effects of dynamical charges that can lead, e.g., to a screening of string tension \cite{Gross1996,Honda2022-PRD}, in part due to the difficulty of solving general LGTs using classical methods \cite{Gattringer2010,Berges2020}.  

In recent years, there is emerging an alternative computing paradigm, quantum simulation, which uses quantum rather than classical hardware to solve quantum many-body problems.
Though in principle universal quantum computers can perform fully scalable quantum simulations, the current devices are still of limited size and coherence times \cite{Hauke2012,Alexeev2021}. 
This state of affairs makes LGTs particularly appealing target models for quantum simulation \cite{Banyuls2020,Alexeev2021}:  
Not only do they host extremely rich physics, the gauge symmetry also presents a new handle on hardware errors as a diagnostic tool and by stimulating the design of new error-mitigation schemes \cite{PRXQuantum,Lamm2020,Kasper2023,Nguyen2022}.  

Here, we experimentally probe geometric confinement in a LGT with $\mathbb{Z}_2$ gauge symmetry (Fig.~\ref{fig:Overview}a) using up to 21 qubits in a superconducting quantum chip.  
We realize the dynamics of the LGT using Trotterization to discretize time and compose the system Hamiltonian through elementary gates (Fig.~\ref{fig:Overview}b). 
The most challenging term, and the heart of our algorithm, is the interaction process between charged matter and gauge fields, which we synthesize as depicted in Fig.~\ref{fig:Overview}c. 
This three-qubit gate uses only 6 native two-qubit gates and (in absence of errors) is fully gauge-invariant.
Efficient circuit synthesis renders single Trotter steps only 8 native two-qubit gates deep and enables us to simulate the dynamics up to 25 Trotter steps, equivalent to a two-qubit gate depth of up to 202.  
In a further experiment, we add the local generators associated with a $\mathrm{U}(1)$ LGT to the model Hamiltonian, permitting us to controllably tune the system from a local $\mathbb{Z}_2$ to a $\mathrm{U}(1)$ gauge symmetry. As we show, the modification of the underlying gauge symmetry can dramatically restrain the charge dynamics. 

Our work adds an important piece to the puzzle of quantum simulations of LGTs with coupled matter and gauge fields.  
At the moment, most existing realizations concern continuous gauge groups \cite{Martinez2016,Bernien2017,Mil2020,Yang2020,Klco2018,Nguyen2022,atas2023simulating}. 
Pioneering laboratory demonstrations for a discrete $\mathbb{Z}_2$ gauge symmetry considered building-blocks \cite{Schweizer2019} and perturbative implementations \cite{Wang2021}. Here, we develop a non-perturbative, in an ideal setting fully gauge-invariant decomposition, enabling us to observe confinement dynamics in a few tens of qubits, to probe long evolution times, and to add an experimentally simple term that tunes the underlying gauge symmetry. 
Our experiments thus open the door towards controlling gauge symmetries in digital quantum simulations and studying the rich dynamics of matter coupled to gauge fields with discrete gauge symmetry.

\textbf{Target theory.---}We are interested here in Abelian LGTs, as sketched in Fig.~\ref{fig:Overview}a. Charged matter lives on sites $i=0,\dots, N-1$ of a one-dimensional spatial lattice. Following a Jordan--Wigner transformation (Methods), we denote the associated operators by Pauli matrices $\sigma_i$. The electric field lives on links between sites, with associated Pauli matrices $\tau^{x}_{i,i+1}$. The target LGT is governed by the Hamiltonian $H=H_{J}+H_{\mathrm{f}}+H_{\mathrm{m}}$, with 
\begin{subequations}
\begin{align}
    \label{equ:h_j}
    H_{J}&=-J\sum_i\left(\sigma_i^+\tau_{i,i+1}^z\sigma_{i+1}^-+\mathrm{h.c.}\right)\,,\\
    \label{equ:h_f}
    H_{\mathrm{f}}&=-f\sum_i\tau_{i,i+1}^x\,,\\
    \label{equ:h_m}
    H_{\mathrm{m}}&=\frac{\mu}{2}\sum_i\left(-1\right)^i\sigma_i^z\,.
\end{align}
\end{subequations}
The most demanding part of the implementation is the three-qubit term $H_{J}$, which couples matter and gauge fields. 
$H_{\mathrm{m}}$ describes the rest mass of matter and $H_{\mathrm{f}}$ is a background field term. This term generates a confining potential, measured by an energy cost proportional to the linear extent of regions with polarized electric field \cite{Schweizer2019,Borla2020, Kebric2023} (Methods). Due to the matter--gauge-field coupling, this linear potential translates over to the charges, mesons are formed and the gauge-invariant fermion correlator decays exponentially as a function of distance \cite{Borla2020,Kebric2021,Kebric2023}. As a result, when $f=0$, the theory is deconfined while charges are confined at $f\neq 0$. To shed further light on confinement in this model, in the Methods we present an analytic derivation of the linear string tension, governed by a complex interplay of kinetic energy of fermions with their rest mass and the background field.

The essence of a LGT is the conservation of a local symmetry. For the case of Hamiltonian $H$, this is a $\mathbb{Z}_2$ gauge symmetry. It is embodied in the conservation of  $G_i^{\mathbb{Z}_2}=-\tau_{i-1,i}^x\sigma_i^z\tau_{i,i+1}^x$, i.e., $[H,G_i^{\mathbb{Z}_2}]=0$, 
$\forall i$. 
Further below, we will demonstrate how an energy penalty term \cite{PRXQuantum} tunes the $\mathbb{Z}_2$ to an approximate $\mathrm{U}(1)$ gauge symmetry, such as governs QED. 
The associated symmetry generator is  $G_i^{\mathrm{U}(1)}=\frac{1}{2}\big[\tau_{i-1,i}^x-\tau_{i,i+1}^x+\sigma_i^z+\left(-1\right)^i\big]$, corresponding to a lattice version of Gauss's law. 
It is these local gauge symmetries that are responsible for the richness of fundamental phenomena in LGTs.   
Conserving these local symmetries for all simulated times $t$ is the central challenge of gauge-theory quantum simulation.

\textbf{Hardware implementation.---}In our implementation, we synthesize the target time-evolution operator generated by $H$ through a first-order Trotter--Suzuki decomposition,
\begin{equation}
    \label{eq:Trotter}
    U(t) = \euler^{-\imunit H t} \simeq \big(\prod_{\ell}\euler^{-\imunit H_\ell\Delta t}\big)^n\,,
\end{equation}
with $n$ the number of Trotter steps, $\Delta t=t/n$ the Trotter time step, and $H_\ell\in\{H_\mathrm{m}, H_\mathrm{f}, H_J\}$.
Taking advantage of a recently elucidated chaos--regular transition in Trotterized dynamics \cite{Heyl2019,Chinni2022}, we can work faithfully at large Trotter steps of $\Delta t\in\{0.2/J,\,0.3/J\}$, and thus reach considerable simulated times within accessible laboratory decoherence times (Methods and Fig.~\ref{fig:trotterization_continuum}). 

We experimentally implement the $\mathbb{Z}_2$ LGT on up to 21 gmon qubits of a superconducting quantum processor of the Sycamore class (Methods). 
This chip natively supports single-qubit $z$-rotations, rotations along arbitrary axes on the $x$-$y$ plane, and two-qubit gates that are close to $\sqrt{\imunit\mathrm{SWAP}}^\dagger=\exp(-\imunit\frac{\pi}{8}\left(\sigma_1^x\sigma_2^x+\sigma_1^y\sigma_2^y\right))$. 
The major implementational challenge is given by the multi-qubit gates constituting $H_J$. 
For these, we have designed an efficient realization using  6 $\sqrt{\imunit\mathrm{SWAP}}^\dagger$ gates and $3$ layers of single-qubit $z$-rotations (Fig.~\ref{fig:Overview}c), the latter of which do not requite execution time (Methods).
To put this into context, decompositions of arbitrary three-qubit circuits require up to 20 CNOT gates \cite{Shende2006}. 
By parallelizing gates acting on neighboring triples, a single Trotter step can be compressed to a depth of 8 layers of $\sqrt{\imunit\mathrm{SWAP}}^\dagger$ gates instead of the 12 one might naively expect (Methods and Fig.~\ref{fig:full_circuit}). 
The respective initial state (Fig.~\ref{fig:Overview}a) is prepared from the fiducial state $\ket{0}^{\otimes L}$, with $L$ being the number of qubits, using single-qubit rotations.   
Compressing the depth of the circuit, altogether a single Trotter step takes 256ns enabling experiments up to 25 Trotter steps for a total run time of an experimental shot up to 6.5$\upmu$s, including state preparation and basis rotations for measurement. 
After the desired number of Trotter steps, we measure the qubits in the computational basis (averaging over 50000 to 200000 shots for each configuration). These measurements hand us a wide range of observables, including matter density, electric-field strength, and degree of violation of $\mathrm{U}(1)$ and $\mathbb{Z}_2$ gauge invariance. 
To increase the data quality (for detailed discussions, see Methods and Fig~\ref{fig:postselection_and_calibration}), we employ Floquet calibration \cite{Arute2020}, which uses periodic sequences to characterize and correct coherent errors in the parameters of entangling gates and their drifts, and average over $5$-$10$ different qubit configurations on the chip. Further, device errors that violate conservation of total matter charge as well as the $\mathbb{Z}_2$ Gauss's law are mitigated by postselecting on these ideally conserved quantities, which is possible thanks to single-site readout.
This postselection also drastically reduces the potential impact of readout errors, as only correlated errors over three nearby qubits would go undetected.

\textbf{Confinement in a \texorpdfstring{$\mathbb{Z}_2$}{Z2} LGT.---}As a first probe of confinement in the target LGT, we initialize the system in the state of Fig.~\ref{fig:Overview}a(i). This state contains a matter defect at the central site and a fully polarized electric field. 
Motion of the charge defect incurs a change in the electric field, which, due to the term $H_{\mathrm{f}}$ requires an energy cost that scales linearly with the traversed distance (i.e., the length of flipped electric field) \cite{Kebric2021}. Thus, in a confined regime, we expect the defect to remain contained to the spatial region where it has been created.
To probe this effect, we quench the system with $H$, with $\Delta t=0.2/J$, $\mu=0$, and $f\in\{0.2J,\, 2.0J\}$, and track in the subsequent dynamics the position of the defect as well as the electric-field excitation through the local $\mathbb{Z}_2$ charge $-\sigma_i^z$ and $\tau_{i,i+1}^x$, respectively.

\begin{figure}[t!]
    \centering
    \includegraphics[width=\columnwidth]{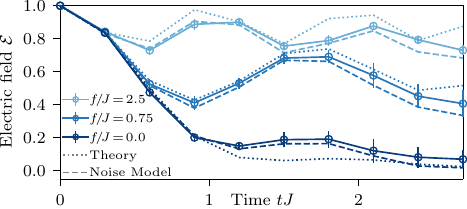}
    \caption{\textbf{Confinement in a $\mathbb{Z}_2$ gauge theory of 16 qubits.} For small strengths of the background field term $f$, charges move through the system and thus flip the electric fields, whose average $\mathcal{E}$ relaxes to 0 (dark curve). As $f$ is increased (lighter curves), the system becomes gradually more confined, preventing a full relaxation of $\mathcal{E}$. Experimental data (symbols connected by solid line) agrees well with ideal theory predictions from exact Trotterized numerical simulations (dotted line), and a microscopic noise model (Methods) can be used to describe remaining deviations (dashed lines).  
    Statistical error bars denote one standard deviation over 10 different circuit assignments around their mean. Data for 8 matter sites plus 8 gauge fields (16 qubits) with periodic boundary conditions, $\mu=0.35J$.}
    \label{fig:z2_dynamics}
\end{figure}

\begin{figure}[t!]
    \centering
    \includegraphics[width=\columnwidth]{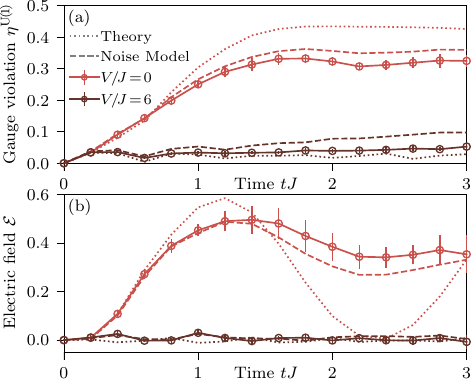}
    \caption{\textbf{Gauge protection and freezing of dynamics induced by modifying the local conservation law.} 
    (a) The dynamics under the $\mathbb{Z}_2$ gauge theory $H$ quickly violates the $\mathrm{U}(1)$ gauge symmetry of the initial state given in Fig.~\ref{fig:Overview}a(iii) ($V/J=0$, light). 
    Adding a penalty term $H_G$ efficiently suppresses $\mathrm{U}(1)$ gauge violations even for rather moderate protection strengths of $V/J=6$ (dark) and controllably tunes the $\mathbb{Z}_2$ into a $\mathrm{U}(1)$ gauge symmetry. 
    (b) The dynamics of the electric field depend drastically on the local conservation law that governs the system. In the original $\mathbb{Z}_2$ theory, $\mathcal{E}(t)$ rises quickly to large values. In stark contrast, in presence of a gauge protection term the emerging local $\mathrm{U}(1)$ symmetry prevents any charges from hopping---the system gets frozen, solely by modifying its local conservation laws.
    Numerical simulations that take a comprehensive noise model (Methods) into account (dashed lines) explain deviations between experiment (symbols connected by solid line) and the ideal theory from exact Trotterized numerical simulations (dotted line), where the leading differences are attributable to parasitic controlled phases appearing with the $\sqrt{\imunit\mathrm{SWAP}}^\dagger$ gates (Methods and Fig.~\ref{fig:u1_protection_noise_comparison}). 
    Statistical error bars denote one standard deviation over 10 different circuit assignments around their mean. Data for 6 matter plus 6 gauge fields (12 qubits) with periodic boundary conditions, $f=\mu=2.5J$.
    }
    \label{fig:u1_protection}
\end{figure}

As the experimental data in Fig.~\ref{fig:Overview}d,e show, for a low value of $f=0.2J$ the action of the term $H_J$ quickly spreads the charge defect. In this parameter range, confinement would make itself felt more significantly only at larger distances than what is probed here.
Enforced by gauge invariance, the charge's motion changes correspondingly the expectation value of the electric field $E_{i,i+1}=\tau_{i,i+1}^x$. 
Upon increasing $f$, the defect remains confined to a narrow region, even up to evolution times considerably larger than the scale of $H_J$. 
The experimental data agrees well with theory; discrepancies between the dynamics observed in the experiments compared to those of the idealised, noise-free theory can be understood by a microscopic noise model (Methods and Figs.~\ref{fig:z2_dynamics},~\ref{fig:u1_protection}).
A small spatial asymmetry arises from the Trotterization of $H_J$ where $\sigma_i^+\tau_{i,i+1}^z\sigma_{i+1}^-+\mathrm{h.c.}$ and $\sigma_{i+1}^+\tau_{i+1,i+2}^z\sigma_{i+2}^- +\mathrm{h.c.}$ are applied subsequently (see Fig.~\ref{fig:Overview}b), and is enhanced by some of the algorithmically chosen circuit assignments (Methods) coincidentally exhibiting higher noise rates asymmetrically towards one end of the system.
For Fig.~\ref{fig:Overview}e, we have chosen to display data up to significantly longer times (as can be seen by the loss of contrast at long times) in order to illustrate the robustness of confinement even in the presence of device errors.
Remarkably, the parameter that is tuned in this experiment couples only to the gauge field, illustrating how charge confinement appears purely as a consequence of coupling between matter and gauge fields. 

To further probe confinement, we initialize the system at half matter filling in the state of Fig.~\ref{fig:Overview}a(ii). 
Figure~\ref{fig:z2_dynamics} displays the dynamics of the average electric field
\begin{equation}
    \mathcal{E}(t)=\frac{1}{N}\sum_i\braket{\tau_{i,i+1}^x(t)}
    \label{equ:electric_field_z2}
\end{equation}
for eight matter sites under periodic boundary conditions plus the eight associated gauge fields. 
Initially, the field is fully polarized. Due to the matter--gauge-field coupling $H_J$, each matter-hopping event flips the electric field in between the involved sites. Thus, under deconfined dynamics the average electric field eventually vanishes.  
In contrast, as $f$ is increased, the charges become confined \cite{Schweizer2019, Borla2020}, and the electric field remains close to its initial configuration. 
This underlying behavior persists to longer time scales and under noise (Methods and Fig.~\ref{fig:z2_dynamics_noisy}).
These findings are indicative of the energy cost associated with changing the polarization of the electric field over large spatial domains.

\textbf{Freezing of dynamics by modifying the gauge constraint.---}Finally, we demonstrate an entirely different mechanism for freezing the charge and electric-field configuration, where not a tuning of the model parameters but a modification of the underlying gauge symmetry drastically restrains the dynamics. 

Consider the apparently simple modification of the gauge-matter coupling from Eq.~\eqref{equ:h_j} to 
\begin{equation}
    H_J^{\mathrm{U}(1)}=-J\sum_i\left[\sigma_i^+(\tau_{i,i+1}^z-\imunit\tau_{i,i+1}^y)\sigma_{i+1}^-+\mathrm{h.c.}\right]\,
\end{equation}
(the term in round brackets corresponds to a spin-raising operator in the $x$ direction of the Bloch sphere). 
The Hamiltonian $H^{\mathrm{U}(1)}=H_J^{\mathrm{U}(1)}+H_{\mathrm{f}}+H_{\mathrm{m}}$ has acquired a $\mathrm{U}(1)$ gauge symmetry, i.e., $[H^{\mathrm{U}(1)},G_i^{\mathrm{U}(1)}]=0$, $\forall i$. 
Now consider the initial state in Fig.~\ref{fig:Overview}a(iii), denoted as $\ket{\psi_{3}}$. This state is consistent with both $\mathbb{Z}_2$ and $\mathrm{U}(1)$ gauge symmetry, with eigenvalues $G_i^{\mathbb{Z}_2}\ket{\psi_{3}}=
(-1)^i\ket{\psi_{3}}$ and 
$G_i^{\mathrm{U}(1)}\ket{\psi_{3}}=
2(-1)^i\ket{\psi_{3}}$. 
If evolved under $H$, only the $\mathbb{Z}_2$ gauge symmetry will be preserved. Moreover, the charges can move by flipping the gauge fields on the involved links. 
In contrast, if $\ket{\psi_{3}}$ is evolved under $H^{\mathrm{U}(1)}$, not only is the $\mathrm{U}(1)$ gauge symmetry preserved, but also the charges are blocked: there are no gauge fields that the action of $H_J^{\mathrm{U}(1)}$ could flip.

To demonstrate this phenomenon, we generate an effective $\mathrm{U}(1)$ gauge symmetry by adding terms proportional to its gauge generators $G_i^{\mathrm{U}(1)}$ \cite{PRXQuantum}. 
Similar symmetry-protecting schemes have been used in the laboratory to perturbatively realize LGTs \cite{Yang2020,Wang2021,Nguyen2022}.
Here, we employ these to tune from the original $\mathbb{Z}_2$ to a $\mathrm{U}(1)$ theory. 

Importantly, the employed term $H_G=V\sum_ic_iG_i^{\mathrm{U}(1)}$ consists of only single qubit operations, which can be absorbed into those already present to implement $H_{\mathrm{f}}$ and $H_{\mathrm{m}}$ (Methods).
The normalized sequence of coefficients $\{c_i\}$ may be chosen \textit{compliant} (Methods) or as simple as an alternating series (Fig.~\ref{fig:u1_protection_extended}).
While $H_G$ commutes with the mass and background-field terms, 
it does compete with the gauge--matter coupling $H_J$. 
Illustratively, one can imagine $H_{G}$ to rapidly rotate the $\mathbb{Z}_2$ gauge field operator $\tau_{i,i+1}^z$ around the $x$-axis \cite{Lamm2020,PRXQuantum,Kasper2023}.  In a time-averaged fashion, $H_J$ thus effectively acquires the $\mathrm{U}(1)$ symmetry of a circle in the $z$-$y$ plane. 
At large $V/J$, one thus approximately obtains $H^{\mathrm{U}(1)}$ \cite{PRXQuantum}. 
Remarkably, we find small values of $V/J=6$ are already sufficient for observing striking effects. 

Initializing the system in $\ket{\psi_3}$ and quenching 
with the $\mathbb{Z}_2$ theory $H$ alone quickly leads to violation of the $\mathrm{U}(1)$ gauge symmetry, quantified through 
\begin{equation}
    \eta^{\mathrm{U}(1)}=\frac{1}{\kappa N}\sum_i\left\langle\left(G_i^{\mathrm{U}(1)}
    -2\left(-1\right)^i\right)^2\right\rangle\,,
    \label{equ:gauge_violation_u1}
\end{equation}  
(light curves in Fig.~\ref{fig:u1_protection}a), where $\kappa=9$ normalizes the theoretical maximum of the gauge violation to unity. 
In contrast, by adding a moderately strong protection term $H_G$, the $\mathrm{U}(1)$ gauge violation is suppressed to small values throughout the entire evolution time (dark curves).
Deviations between experimental data and noise-free simulations, which can be explained by a comprehensive noise model (Methods), are dominated by parasitic controlled phases appearing with the $\sqrt{\imunit\mathrm{SWAP}}^\dagger$ gates (Methods and Fig.~\ref{fig:u1_protection_noise_comparison}).

This modification of the underlying gauge symmetry has dramatic repercussions on the entire system dynamics: In the $\mathbb{Z}_2$ theory (small $V/J$), the charges are free to move and we observe a strong dynamics also in the electric field (light curves in Fig.~\ref{fig:u1_protection}b).
Instead, upon tuning the symmetry towards $\mathrm{U}(1)$ by means of $H_G$, charges become frozen and the electric field remains close to its initial value for the entire evolution time.

\textbf{Outlook.---}Our work opens the door to studying a wide range of  phenomena in LGTs of coupled matter and gauge fields with discrete symmetries. Ideal target scenarios are, e.g., exotic out-of-equilibrium phenomena such as quantum many-body scars \cite{Aramthottil2022} and disorder-free localization \cite{Disorder-Free_Localization_2}. 
As the cited numerical studies suggest, the relevant features of these effects appear already at system sizes of $L=4-18$ and times around $Jt=10$. Increases in circuit depths of only a factor of $2-4$ may thus put these phenomena within reach of the capabilities presented here.
In this effort towards larger coherence times, strongly-correlated systems given by gauge theories provide an ideal testbed for advanced error-mitigation strategies, such as probabilistic error cancellation \cite{Temme2017}, zero-noise extrapolation \cite{Li2017, Temme2017}, readout-error mitigation \cite{Maciejewski2020, Bravyi2021}, learning-based mitigation \cite{Czarnik2021, strikis2021learning-based}, purity constraints \cite{huggins2021virtual, obrien2023purification-based}, or approaches to mitigate parasitic gates \cite{Lao_2022}.
Given the large computational challenges when attempting to connect early-time behavior to late-time dynamics, thermalization of gauge theories is another outstanding question with large relevance, e.g., to high-energy experiments such as heavy-ion collisions \cite{Berges2020}. A first step towards illuminating this type of question has been taken recently in an optical lattice setup simulating a $\mathrm{U}(1)$ gauge theory, where few gauge-invariant hopping events were sufficient to already see main features of thermalization \cite{Zhou2022}.

The necessity of precisely engineering multi-qubit interactions and local constraints is currently imparting a strong momentum onto the field, as it pushes devices beyond their current limits and drives the development of new error mitigation protocols \cite{Yang2020,Lamm2020,Kasper2023,Wang2021,Nguyen2022}, such as the one implemented here \cite{PRXQuantum}. 
Promising proposals exist for generalizing such schemes to more complex situations, including non-Abelian gauge theories \cite{Non-Abelian,Lamm2020,Kasper2023}. 

Such investigations may also develop interesting cross-fertilizations to other fields. E.g., postselecting on the Gauss' law is similar to syndrome measurements in error-correcting codes \cite{Young2013,gottesman1997stabilizer}. 
Beyond the immediate context of LGTs, three-body gates as designed here or similar (Methods) are a key ingredient for manifold physical scenarios, including cluster Ising models \cite{Decker2020}, topological phase transitions \cite{Smith2022}, spin chirality \cite{Wang2019}, multipartite entanglement \cite{Wang2019}, and even non-stoquastic and parity quantum annealing \cite{Hauke2020}.

\textbf{Acknowledgements.---}We are grateful to the Google Quantum AI team, in particular Eric Ostby and Markus Hoffmann, for support and discussions. We also thank Haifeng Lang for discussions. 
We acknowledge participation in the Google Quantum AI Early Access Program, within which all quantum computations have been performed remotely between two continents, employing Cirq \cite{cirq}. Further numerical simulations have been performed with the help of qsim \cite{qsim}.
This project has received funding from the European Research Council (ERC) under the European Union’s Horizon 2020 research and innovation programme (grant agreement No 804305, StrEnQTh, and grant agreement No 948141, SimUcQuam), within the QuantERA II Programme from the European Union’s Horizon 2020 research and innovation programme (grant agreement No 101017733), from the European Union’s Horizon Europe research and innovation programme (grant agreement 101080086, NeQST), the Deutsche Forschungsgemeinschaft (DFG, German Research Foundation) under Germany's Excellence Strategy (EXC-2111 -- 390814868), the Italian Ministry of University and Research (MUR) through the FARE grant for the project DAVNE (Grant R20PEX7Y3A), the Google Research Scholar Award ProGauge, Provincia Autonoma di Trento, and Q@TN, the joint lab between University of Trento, FBK-Fondazione Bruno Kessler, INFN-National Institute for Nuclear Physics, and CNR-National Research Council.

\bibliography{bibliography.bib}

\clearpage

\counterwithin*{figure}{section}
\renewcommand{\thefigure}{S\arabic{figure}}
\setcounter{figure}{0}

\section*{Methods}

\subsection{Target lattice gauge theories (LGTs)}

\subsubsection{\texorpdfstring{$\mathbb{Z}_2$}{Z2} LGT}

The $\mathbb{Z}_2$ LGT in 1+1D considered in the main text is equivalent to the more common formulation with fermionic matter, described by the Hamiltonian \cite{Schweizer2019,Kebric2021} 
\begin{align}
\label{equ:HZ2}
H^{\mathbb{Z}_2}=&-J\sum_i\left(\psi_i^\dagger \tau_{i,i+1}^z \psi_{i+1}+\mathrm{h.c.}\right)\\
&-f\sum_i\tau_{i,i+1}^x\nonumber\\
&+\mu\sum_i\left(-1\right)^i\psi_i^\dagger \psi_i\nonumber\,.
\end{align}
Here, $\psi_i^\dagger$ ($\psi_{i}$) are fermionic creation (annihilation) operators and---compared to works such as Refs.~\cite{Schweizer2019,Kebric2021}---we have added a staggered fermion mass term $\sim \mu$. This term is compatible with the symmetries of the system and enables us to connect to the $\mathrm{U}(1)$ theory discussed below, where it is commonly included. Physically, it can be interpreted as giving a positive (negative) rest energy to (anti-)particles on even (odd) matter sites.  

In the fermionic formulation of the $\mathbb{Z}_2$ LGT, the symmetry generators are given by
\begin{equation}
G_i^{\mathbb{Z}_2}=\euler^{\imunit\pi \psi_i^\dagger \psi_i}\prod_{j:\langle i,j\rangle}\tau_{\langle i,j\rangle}^x\,.
\label{equ:generators_z2}
\end{equation}
Here, $j:\langle i,j\rangle$ denotes the product over all links connected to matter site $i$. In our 1+1D case, these are $\tau_{i,i-1}^x$ and $\tau_{i,i+1}^x$. Gauge symmetry is equivalent to $[H^{\mathbb{Z}_2},G_i^{\mathbb{Z}_2}]=0\ \forall i$, and the conservation of the eigenvalues $g_i^{\mathbb{Z}_2}=\pm1$ of the generators, $G_i^{\mathbb{Z}_2}\ket{\psi(t)}=g_i^{\mathbb{Z}_2}\ket{\psi(t)}$ $\forall t$.

The gauge violation respective to an initial, gauge-invariant state with eigenvalues $g_i^\mathrm{in}$ of the generators in Eq.~\eqref{equ:generators_z2} is quantified by
\begin{equation}
    \eta^{\mathbb{Z}_2}=\frac{1}{N\kappa^\prime}\sum_i\left\langle\left(G_i^{\mathbb{Z}_2}-g_i^\mathrm{in}\right)^2\right\rangle\,,
    \label{equ:gauge_violation_z2_general}
\end{equation}
where $\kappa^\prime$ normalizes the maximal possible gauge violation to unity.
This quantity can be understood as an average distance in Hilbert space from the target superselection sector defined by $\{g_i^\mathrm{in}\}$ \cite{Decoherence}, and non-zero values are equivalent to the presence of gauge errors. 

Substituting the fermionic degrees of freedom by spins-$1/2$ via a Jordan--Wigner transformation, 
\begin{align}
    \label{equ:jordan_wigner}
    \psi_i^\dagger&=\euler^{+\imunit\pi\sum_{k<i}\sigma_k^+\sigma_k^-}\sigma_i^+\nonumber\\
    \psi_i&=\euler^{-\imunit\pi\sum_{k<i}\sigma_k^+\sigma_k^-}\sigma_i^-
\end{align}
with $\sigma^\pm=(\sigma^x\pm\imunit\sigma^y)/2$, which retains the canonical fermion anti-commutation relations,  
yields the Hamiltonian $H$ and gauge generators reported in the main text.

\subsubsection{\texorpdfstring{$\mathrm{U}(1)$}{U(1)} LGT}

Consider the Kogut--Susskind Hamiltonian formulation of QED in 1+1D with staggered fermions \cite{Schwinger1962,Kogut1975}, 
\begin{align}
\label{equ:hamiltonian_u1_initial}
H^{\mathrm{U}(1)}=&-2J\sum_i\left(\psi_i^\dagger U_{i,i+1}\psi_{i+1}+\mathrm{h.c.}\right)\nonumber\\
&+\mu\sum_i\left(-1\right)^i\psi_i^\dagger\psi_i+\frac{g^2}{2}\sum_i\left(E_{i,i+1}+E_0\right)^2\,,
\end{align}
where the role of $J$ and $\mu$ is as in $H^{\mathbb{Z}_2}$. The last term gives the electric field energy and $E_0$ is a background field, which in this theory is equivalent to a topological $\theta$ angle \cite{Surace2020,Zache2022}. The electric field $E_{i,i+1}$ and parallel transporters $U_{j,j+1}$ fulfil the commutation relation
\begin{equation}
\label{equ:commutator_u1}
    [E_{i,i+1},U_{j,j+1}]=\delta_{ij}U_{i,i+1}\,.
\end{equation}
This model has a continuous $\mathrm{U}(1)$ gauge symmetry with generators 
\begin{equation}
    G_i^{\mathrm{U}(1)}=E_{i-1,i}-E_{i,i+1}+\psi_i^\dagger\psi_i+\frac{\left(-1\right)^i-1}{2}\,.
    \label{equ:generators_u1}
\end{equation}
We have $[H^{\mathrm{U}(1)},G_i^{\mathrm{U}(1)}]=0\ \forall i$, and the eigenvalues $g_i^{\mathrm{U}(1)}$ of the generators, $G_i^{\mathrm{U}(1)}\ket{\psi(t)}=g_i^{\mathrm{U}(1)}\ket{\psi(t)}$, are conserved under Eq.~\eqref{equ:hamiltonian_u1_initial}.

The gauge fields can be expressed within the quantum link model formalism \cite{Wiese2013} with $S=1/2$ as $U_{i,i+1}=\tau_{i,i+1}^+$, $E_{i,i+1}=\frac{1}{2}\tau_{i,i+1}^z$, which retains the system's symmetries and the commutation relation of Eq.~\eqref{equ:commutator_u1}. Again applying the Jordan--Wigner transformation as in Eq.~\eqref{equ:jordan_wigner} yields
\begin{align}
    \label{eq:HU1_QLM}
    H^{\mathrm{U}(1)}=&-2J\sum_i\left(\sigma_i^+\tau_{i,i+1}^+\sigma_{i+1}^-+\mathrm{h.c.}\right)\nonumber\\
    &+\frac{\mu}{2}\sum_i\left(-1\right)^i\sigma_i^z+\frac{E_0g^2}{2}\sum_i\tau_{i,i+1}^z\,.
\end{align}
Here, we employ $(\tau_{i,i+1}^z)^2=\boldone$ and neglect constant terms.
Performing a basis rotation on the gauge fields such that $\tau_{i,i+1}^x\leftrightarrow\tau_{i,i+1}^z$, $\tau_{i,i+1}^y\rightarrow-\tau_{i,i+1}^y$, thus $\tau_{i,i+1}^+\rightarrow (\tau_{i,i+1}^z-\imunit\tau_{i,i+1}^y)/2$, and redefining the constant $E_0g^2/2\rightarrow-f$ yields the $\mathrm{U}(1)$ gauge theory and generators reported in the main text.

The gauge violation respective to an initial, gauge-invariant state with eigenvalues $g_i^\mathrm{in}$ of the generators in Eq.~\eqref{equ:generators_u1} is quantified by
\begin{equation}
    \eta^{\mathrm{U}(1)}=\frac{1}{N\kappa}\sum_i\left\langle\left(G_i^{\mathrm{U}(1)}-g_i^\mathrm{in}\right)^2\right\rangle\,,
\end{equation}  
with $\kappa$ normalizing the maximal possible gauge violation to unity.

\subsubsection{Confinement}

The question when confinement in interacting gauge theories with dynamical matter emerges is highly non-trivial. E.g., in the Schwinger model, corresponding to the continuum limit of the Hamiltonian in Eq.~\eqref{equ:hamiltonian_u1_initial}, dynamical fermions can screen the string tension between two probe particles and thus suppress confinement \cite{coleman1975charge,Gross1996,Honda2022-PRD}. Specifically, within mass perturbation theory, the string tension for probe particles with charge $qg$ is proportional to $\mu[1-\cos(2\pi q)]$, i.e., it vanishes for vanishing rest mass of the dynamical fermions as well as for any integer probe charge. The reason behind the latter behavior is the periodicity of the vacuum under rotations of the topological $\theta$-angle by $2\pi$. A $\theta$-angle corresponds to a background field [see Eq.~\eqref{equ:hamiltonian_u1_initial}], which can be thought of as being generated by a boundary charge of $g \theta /(2\pi)$. Now, moving a pair of probe particles all the way to the edge of the system provides an additional boundary charge of $gq$. For integer $q$, this charge can be absorbed by a rotation of the $\theta$-angle by multiples of $2\pi$. As the ground-state energy is invariant under such a rotation, it costs no energy to move an integer probe charge to infinity, resulting in a vanishing string tension. This argument no longer holds once the periodicity of the ground state under addition of the background field is lost, lifting this fundamental difference between integer and non-integer probe charges. That happens in the QLM for $S<\infty$ [see also Eq.~\eqref{eq:HU1_QLM}] as well as in the $\mathbb{Z}_2$ LGT considered in this work.

For the $\mathbb{Z}_2$ LGT at hand [Eq.~\eqref{equ:HZ2}], it has been argued that the theory is confining as soon as $f\neq 0$, as illustrated, e.g., through an exponential decay of fermion correlators with distance or a change of the spatial periodicity of Friedel oscillations \cite{Borla2020,Kebric2021,Kebric2023}. This can be understood to occur thanks to a linear potential generated by the term $H_f$, as has been argued for isolated fermion dimers in the limit of $f\to\infty$ in Refs.~\cite{Schweizer2019,Kebric2023}. This scenario corresponds to the one in Fig.~\ref{fig:Overview} of the main text, where the length of the matter dimer corresponds to the distance of the dynamical defect from a static background charge at the center of the chain where the Gauss's law has eigenvalue switched to $g=+1$. In what follows, we want to refine this discussion for the sector with fermionic filling of one-half. 

Consider the ground state of $H$ as in Eq.~\eqref{equ:HZ2} with open boundary conditions and a background field of $\tau_\mathrm{bg}^x=1$ entering the system. We choose the sector with fermionic filling of one-half and $\mathbb{Z}_2$ Gauss' law $g^{\mathbb{Z}_2}_i=(-1)^i$. In the limits of $\mu/J\gg 1$ or $f/J\gg 1$, the ground state has a fully polarized field threading through the system, $\ket{\psi_0}=\ket{{\color{gray}\rightarrow}  \bullet \rightarrow \circ \rightarrow  \bullet \rightarrow \circ \cdots \rightarrow  \bullet \rightarrow \circ {\color{gray}\rightarrow}}$ (here, $\ket{\rightarrow/\leftarrow}_{i,i+1}$ denotes the plus/minus eigenstate of $\tau_{i,i+1}^x$, and $\ket{\bullet/\circ}_{i}$ a filled/empty matter site; grey color denotes boundary fields entering the system of interest from the outside). 
For a system of $N$ matter sites and $N-1$ links in between, within second-order perturbation theory in $J$ this state has an energy of 
$\varepsilon_0=(N-1)(-f -\frac{\mu}{2} - \frac{J^2}{2f+2\mu})-\frac{\mu}{2}$.

We can now test the energy of a pair of charges at various distances added to this state. For simplicity, we add a static pair of charges at the boundaries just outside the system, equivalent to flipping the state of the electric field entering the system. 
In the limit of $\mu\gg f$ and $\mu\gg J$, the new ground state obeying the Gauss' law $g_i=(-1)^i$ is $\ket{\psi_1}=\ket{{\color{gray}\leftarrow}  \bullet \leftarrow \circ \leftarrow  \bullet \leftarrow \circ \cdots \leftarrow  \bullet \leftarrow \circ {\color{gray}\leftarrow}}$, i.e., the charges remain at their staggered ordering as before and a flip of the electric field compensates for the background field entering the system. 
Within second-order perturbation theory in $J$, the energy of this state is 
$\varepsilon_1=(N-1)(f -\frac{\mu}{2} - \frac{J^2}{2\mu-2f})-\frac{\mu}{2}$. 
If we are instead in the limit $f\gg \mu$ and $f \gg J$, it is energetically favourable for the charges to completely overscreen the background field, leading to the state $\ket{\psi_1'}=\ket{{\color{gray}\leftarrow}  \circ \rightarrow \bullet \rightarrow  \circ \rightarrow \bullet \cdots \rightarrow  \circ \rightarrow \bullet {\color{gray}\leftarrow}}$. 
The energy of this state is 
$\varepsilon_1'=(N-1)(-f +\frac{\mu}{2} - \frac{J^2}{2f-2\mu})+\frac{\mu}{2}$.

Thus, in the limit of small hopping \footnote{In a gauge theory with a well-defined continuum limit, the hopping is related to the lattice spacing $a$ as $J=1/{2a}$, i.e., the perturbative arguments may break down as one approaches the continuum limit $a\to 0$. Here, we are purely interested in the behavior of the lattice theory.} 
the energy of a charge pair at distance $N$ on top of the ground state is 
\begin{align}
\varepsilon_1-\varepsilon_0&=(N-1)2f\left(1 - \frac{J^2/2}{\mu^2-f^2}\right) \qquad\qquad \mu\gg J,f\,, \\
\varepsilon_1'-\varepsilon_0&=(N-1)\mu\left(1 - \frac{J^2}{f^2-\mu^2}\right)+\mu \qquad f\gg J,\mu\,.     
\end{align}
That is, the energy cost associated to the creation of a charge pair increases linearly with distance. 
In the considered sector of gauge symmetry and particle number, the resulting string tensions are $2f\left(1 - \frac{J^2/2}{\mu^2-f^2}\right)$ ($\mu\gg J,f$) and $2\mu \left(1 - \frac{J^2/2}{f^2-\mu^2}\right)$ ($f\gg J,\mu$), respectively. 
Interestingly, they disappear if the subdominating parameter of either $\mu$ or $f$ vanishes. 
They are also very different from the one expected from a classical Coulomb law, $\sigma_{\mathrm{Coulomb}}=g^2 q^2 /2$ (where for us $q=1$), in particular due to the non-trivial dependence on all the parameters of the Hamiltonian, indicating the complex quantum many-body nature of the string tension in this model.

In the main text, we are interested in illustrating how this confining linear potential becomes manifest in the out-of-equilibrium dynamics of the model. We see, e.g., in Figs.~\ref{fig:Overview} and \ref{fig:z2_dynamics} how the large energy cost associated with flipping the electric field from its polarized initial value limits the spread of probe charges and pins the electric field to its initial value. 
It will also be interesting to extract directly the string tension of the model. For that, one can measure the energy cost of the creation of probe charges above the ground-state energy. This can be achieved in quantum simulators using adiabatic state preparation, as has been discussed, e.g., in the context of the Schwinger model in Refs.~\cite{Honda2022-PRD,Honda2022-PTEP}. 
In particular, such approaches may permit to extend the above discussion into the challenging non-perturbative regimes where $J$, $\mu$, and $f$ compete.\\

\subsection{Hardware}
\label{sec:hardware}

We implement the target LGTs in superconducting quantum chips of the Sycamore class, as have been described, e.g., in Refs.~\cite{Arute2019,Arute2020}. 
The layout of these chips is that of a square lattice, which permits us to naturally map the qubits to the degrees of freedom of the 1+1D LGT by alternating assignment of matter sites and gauge links. 
More specifically, we employ the Rainbow and Weber chips, which host $23$ and $53$ qubits, respectively, of which we use up to 16 qubits under periodic and 21 qubits under open boundary conditions. The experiments were performed remotely via the cloud on devices that were calibrated using automated calibration procedures. 

The Sycamore-class quantum chips natively implement arbitrary single qubit rotations around the $z$ axis (which can be implemented virtually without execution time) and around arbitrary axes in the $x$\nobreakdash--$y$ plane. The latter take a gate time of $25\mathrm{ns}$ and their average error rate per gate is about $0.1\%$. 
In addition, these chips have a native two-qubit gate that approximates the form $\exp(-\imunit\frac{\theta}{2}\left(\sigma_1^x\sigma_2^x+\sigma_1^y\sigma_2^y\right))\exp(-\imunit\frac{\phi}{4}\left(\boldone-\sigma_1^z\right)\left(\boldone-\sigma_2^z\right))$. 
Since $\theta$ is close to $\pi/4$ and $\phi\simeq0.138\pm0.015$ \cite{Arute2020}, this gate can to leading order be approximated as a $\sqrt{\imunit\mathrm{SWAP}}^\dagger$ gate, with average error rates per gate of 1\nobreakdash--1.5\%. In some situations, though, the parasitic C-phase term $\propto\phi$ can have considerable effects on the simulated dynamics \cite{Yan2018, Foxen2020, Arute2020, Arute2020-hartree, tan2023realizing}, and in our experiments it can spoil gauge invariance.
In most of the present scenarios, however, we found that postselection onto the gauge-invariant and charge-conserving subspace yielded satisfactory suppression of this error type. For the scenario of Fig.~\ref{fig:u1_protection} and particularly for $V/J=0$, the numerical simulations of the parasitic C-phase term displayed in Fig.~\ref{fig:u1_protection_noise_comparison}, which are projected into the gauge-invariant and total matter charge conserving subspace at readout as usual, demonstrate that it can also cause unwanted deviations in the dynamics within this subspace.
In the future, it may be interesting to test in how far the attainable simulation times can be improved by using strategies to mitigate these types of parasitic terms \cite{Lao_2022}. Here, we chose keeping the 3-qubit gate as compact as possible, in order to avoid additional error terms that a re-decomposition taking the parasitic phase into account would incur by requiring a longer gate sequence.

A further relevant error term stems from readout, whose median values over the chips are on the order of $1\%$ ($\ket{0}$-state) to $5\%$ ($\ket{1}$-state), see also Fig.~\ref{fig:noise_histograms}. Postselection onto the gauge-invariant subspace proves highly beneficial also for these types of errors, since in order to go undetected they would need to appear simultaneously over three neighbouring qubits that are involved in two overlapping given Gauss' law generators. Such correlated readout errors are suppressed by the third power.
To give an example of this benefit of postselection, for the data in Fig.~\ref{fig:z2_dynamics} at the initial time (directly after preparation) the deviation from the ideal value is only $1-E(t=0) = 0.0000234$, much smaller than the individual readout error.

A more comprehensive description of the hardware used can be found in Ref.~\cite{Arute2019} and its Supplementary Material. Section~\ref{sec:noise_calculations} provides more details on error types present in this hardware.

\subsection{Additional details on the quantum circuit}

\begin{figure*}[t!]
    \centering
    \includegraphics[width=\textwidth]{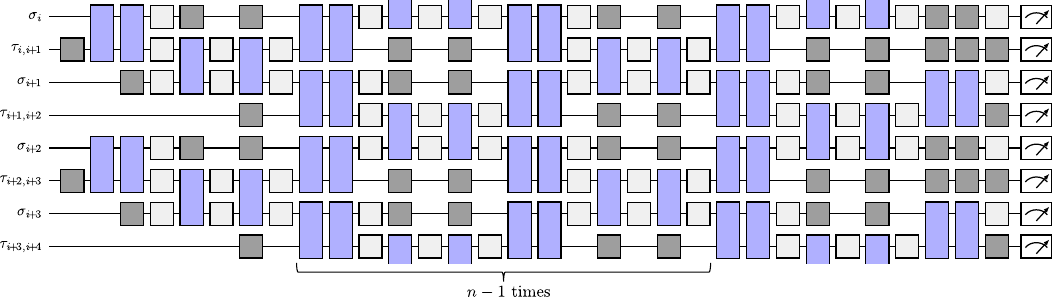}
    \caption{\textbf{Circuit diagram for $n$ Trotter steps.} Two-qubit $\sqrt{\imunit\mathrm{SWAP}}$ gates (blue) are executed in parallel to generalized single-qubit rotations (dark gray). Most of the latter arise from the purpose of avoiding idling qubits by acting as spin echo sequences, hence reducing errors induced by crosstalk with neighbouring, active qubits. Layers consisting only of $z$-rotations (light gray) require zero execution time. At the beginning of the circuit, it is beneficial to delay the initialisation of qubits as far as possible. Shown for an excerpt of eight qubits and the initial states of either Fig.~\ref{fig:Overview}a(ii) or (iii). 
    }
    \label{fig:full_circuit}
\end{figure*}

Since the single qubit rotations in Fig.~\ref{fig:Overview}b,c can be naturally realized in the Sycamore chip, the key implementation challenge is the decomposition of the matter--gauge coupling into single- and two-qubit gates.
This decomposition can be realized as
\begin{align}
\label{Uijk}
	U_{ijk}(\alpha)=&\euler^{-\imunit\alpha\left(\sigma_i^+\tau_{j}^z\sigma_k^-+\mathrm{h.c.}\right)} \nonumber\\
	=&\euler^{\imunit\pi}\sqrt{\imunit\mathrm{S}}_{ij}\sqrt{\imunit\mathrm{S}}_{ij}R_i^z(\pi)R_j^z(-\pi/4)R_k^z(-\pi/4)\nonumber\\
	&\sqrt{\imunit\mathrm{S}}_{jk} R_j^z(\pi-\alpha)R_k^z(\alpha)\sqrt{\imunit\mathrm{S}}_{jk}\nonumber\\
	&R_j^z(\pi/4)R_k^z(\pi/4)\sqrt{\imunit\mathrm{S}}_{ij}\sqrt{\imunit\mathrm{S}}_{ij}\,,
\end{align}
where $i$, $k$ label qubits representing matter sites connected via qubit $j$ representing the linking gauge field, and $\sqrt{\imunit\mathrm{S}}_{ij}$ is denoting the square root of iSWAP gate ($\sqrt{\imunit\mathrm{SWAP}}_{ij}$). It is useful to note that the latter is equal to the native $\sqrt{\imunit\mathrm{S}}_{ij}^\dagger$ up to $z$-rotations as $\sqrt{\imunit\mathrm{S}}_{ij}^\dagger=\mathrm{R}_i^z\left(\pi/2\right)\mathrm{R}_j^z\left(-\pi/2\right)\sqrt{\imunit\mathrm{S}}_{ij}\mathrm{R}_i^z\left(-\pi/2\right)\mathrm{R}_j^z\left(\pi/2\right)$.

The generator of an arbitrary 3-qubit unitary such as $U_{ijk}$ can be decomposed into products of three Pauli matrices acting on the three involved qubits (plus products of lesser numbers of operators). As the present 3-qubit gate has a specific physical interpretation (gauge-invariant hopping), its generator is a Hamiltonian consisting of only two 3-qubit operators (XZX and YZY) whose strength is equal.
In addition, this particular 3-qubit gate can be interpreted in terms of fermionic basis transformations. The two-qubit fermionic simulation gates, i.e., the broader family of native two-qubit gates for arbitrary $\theta,\phi$ of section~\ref{sec:hardware}, and single-qubit $z$-rotations can be mapped as two- and one-mode single-particle fermionic basis transformations via a Jordan--Wigner transformation and together allow for the construction of multi-qubit gates corresponding to arbitrary single-particle fermionic basis transformations, including the 3-qubit gate under consideration, effectively reducing the target unitary dimension.
This large structure of the generator aids the identification of the above efficient decomposition and makes the implementation of the corresponding unitary feasible on current devices. Similarly efficient decompositions can be found for other native gate sets (e.g., 6 CNots plus single-qubit rotations suffice to decompose the gauge-invariant hopping) 
or other 3-qubit gates with a comparable level of structure in their generator. E.g., gates generated by $\sigma_i^z\sigma_{j}^z\sigma_k^z$, which are relevant for various applications \cite{Wang2019,Decker2020,Hauke2020,Smith2022}, can be implemented with either 4 CZ gates or 6 $\sqrt{\mathrm{iSWAP}}$ gates.

In a concrete implementation, to limit decoherence, it is desirable to keep the execution time of the matter--gauge coupling as short as possible. Furthermore, idling qubits should be avoided as these are subject to erroneous $z$-rotations due to crosstalk and low-frequency noises. To account for these, in our implementation we compress the circuit by executing neighbouring gates in parallel wherever possible. Beyond this, we add spin echo sequences by randomly chosen angles on idling qubits. A sketch of the full gate sequence can be found in Fig.~\ref{fig:full_circuit}.

The theoretical predictions used for comparison in the figures result from numerically exact propagation of the many-body wave function through these quantum circuits.

\subsection{Use of large Trotter step sizes}

In principle, we are interested in the simulation of the LGT dynamics in continuous time. As recent works have shown \cite{Heyl2019,Sieberer2019,Kargi2021,Chinni2022}, this does not require taking the limit of vanishing Trotter steps $\Delta t\to 0$ in Eq.~\eqref{eq:Trotter} as one might expect from worst-case bounds \cite{Lloyd1996}. Instead, viewing the Trotterized dynamics as a periodic sequence of pulses, one can use techniques from periodically driven systems to identify a sharp crossover between regular and chaotic quantum dynamics. In the regular regime, the system dynamics deviates only perturbatively from the dynamics at $\Delta t\to 0$. Numerical benchmarks on finite systems show that the regular regime extends to surprisingly large Trotter step sizes and that its effect holds up to essentially infinite times \cite{Heyl2019,Sieberer2019,Kargi2021,Chinni2022}. This enables a controlled simulation of continuous-time dynamics with large $\Delta t$. 

\begin{figure*}[t!]
    \centering
    \includegraphics[width=\textwidth]{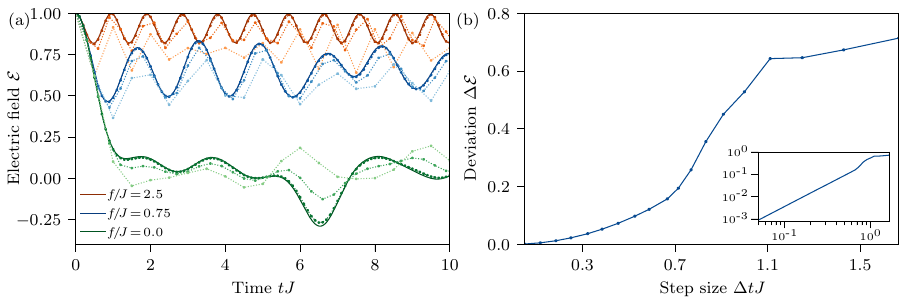}
    \caption{\textbf{Robustness of results against Trotter step size, numerical simulations.} (a) The range of Trotter step sizes used in our experiments introduces only small deviations of local observables from the continuum-time results (solid lines), illustrated here for the scenario of main-text Fig.~\ref{fig:z2_dynamics}, with step sizes $0.1/J$, $0.3/J$, $0.5/J$ (dotted lines, shading from light to dark). (b) For small Trotter step sizes $\Delta t$, the deviation of the electric field from the time-continuum result increases quadratically until a sharp crossover from this perturbative to a chaotic regime around $\Delta t\approx 0.7/J$, see inset for log-log scale (example for the scenario of panel (a) with $f/J=0.75$). 
    One can thus find an optimal balance of choosing $\Delta t$ below the threshold, ensuring reliable results with controllable errors, while keeping it relatively large in order to reach significant evolution times at fixed number of Trotter steps (in our experiments, we use $\Delta t\in\{0.2/J,\,0.3/J\}$). 
    }
    \label{fig:trotterization_continuum}
\end{figure*}

In Fig.~\ref{fig:trotterization_continuum}, we numerically investigate this effect for the $\mathbb{Z}_2$ LGT considered here. 
Panel (a) displays the time evolution of the electric field $\mathcal{E}$, which quickly approaches the ideal, continuum-time curve ($\mathcal{E}_0=\lim_{\Delta t\to 0}\mathcal{E}$) as the Trotter step is diminshed. Panel (b) displays the medium-time average of the deviation of the electric field, defined as $\Delta\mathcal{E}=\frac{1}{m}\sum_{k=1}^{m}\left|\mathcal{E}_{\Delta t}\left(k\Delta t\right)-\mathcal{E}_0\left(k\Delta t\right)\right|$ with $m=t_f/\Delta t$ the number of steps for the particular Trotter step size $\Delta t$ and $t_f=10/J$ the final evolution time. At small $\Delta t$, the deviation increases only perturbatively as $(\Delta t)^2$ (see also \cite{Heyl2019}).    
As this additional data shows, even large Trotter steps of $\Delta t\approx 0.5/J$ and above lie in the perturbative, regular regime. In our quantum simulations, we use $\Delta t\in\{0.2/J,\,0.3/J\}$, enabling us to reach significant simulation times $t=n \Delta t $ at given number of Trotter steps $n$ while keeping the Trotterization error at a controlled level. 

\subsection{Qubit selection}

For each studied scenario, we average over several configurations of qubits (chains or rings of given length) in order to reduce errors.  
Since the performance of qubits and gates varies in position as well as time, it is crucial to identify those that perform well in a given experimental session.  
While calibration data for readout, single-, and two-qubit gates can indicate qubits that perform especially bad, it does not pick out the best ensembles of qubits. Hence, we use a layered approach. At the beginning of a session, we use the calibration data to exclude the qubits and two-qubit gates identified to have particularly high error rates. In order to reduce the number of possible configurations (i.e., chains or rings for our purposes) to an amount feasible to benchmark, we further employ a simple cost function using the calibration data as input. We then use a minimal model for a specified parameter set to pick suitable ensembles among the remaining possible sets of qubits. To address time-dependent drifts within the duration of a session, we repeat the minimal model test between experiments, restricted to qubit sets performing well in the test at the beginning of the session.
With this procedure, we pick $5$ chains of qubits in the case of Fig.~\ref{fig:Overview}d,e and $10$ rings with periodic boundary conditions for Fig.~\ref{fig:z2_dynamics} and Fig.~\ref{fig:u1_protection}, on which the experimental runs are executed and averaged over.
Error bars in the figures denote one standard deviation over the results for different circuit assignments. In scenarios where the ideal dynamics display large oscillations, small errors can lead to sizable deviations between different realizations and, thus, growing error bars. Instead, if the ideal dynamics prescribes only small oscillations around a constant or follows a damped decay, small differences between realizations play a less prominent role.

\subsection{Postselection}
\label{sec:postselection}

The local gauge symmetry also proves a useful tool to mitigate errors: any errors that break this local conservation law, akin to a stabilizer used in quantum error correction \cite{Young2013,gottesman1997stabilizer}, can be identified through postselection. This permits us to considerably improve the quality of the simulation. 

As the implemented circuits respect the $\mathbb{Z}_2$ gauge symmetry by design, any experimental shots where the eigenvalues of the generators Eq.~\eqref{equ:generators_z2} differ from those of the initial state are known not to be faithful. This is equivalent to the gauge-violation in Eq.~\eqref{equ:gauge_violation_z2_general} being non-zero. 
In addition, the total $\mathbb{Z}_2$ charge $\sum_i\sigma_i^z$ should ideally be conserved in our implementations, which we can use in postselection to further eliminate faulty experimental runs.

\begin{figure*}[t!]
    \centering
        \includegraphics[width=\textwidth]{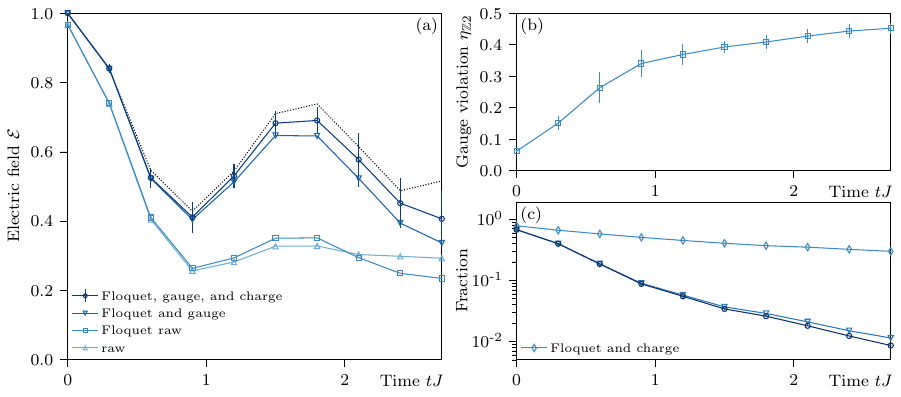}
    \caption{\textbf{Postselection and Floquet calibration.} (a) Raw data (lightest curve) display sizable discrepancies to the theoretical predictions from exact Trotterized numerical simulations (dotted). Floquet calibration yields slight improvements, while postselection on local $\mathbb{Z}_2$ gauge symmetry significantly enhances the data quality. Poststelection on global $\mathbb{Z}_2$ charge results in a further small improvement. Data shown in the main text includes all these steps, illustrated here on data of Fig.~\ref{fig:z2_dynamics} for 8 matter sites plus 8 gauge fields (16 qubits) with periodic boundary conditions and $f/J=0.75$; statistical error bars denote one standard deviation over 10 different circuit assignments around their mean (shown only for one data set for clarity; error bars are of similar magnitude for other data sets). (b) Dynamics of $\mathbb{Z}_2$ gauge violation of raw data with Floquet calibration. (c) Fraction of states fulfilling either of the two or both postselection criteria. The steeper decrease of the fraction of gauge-symmetric states illustrates that this is a stronger postselection criterion.}
    \label{fig:postselection_and_calibration}
\end{figure*}

While total charge conservation is commonly used as a postselection criterion in quantum simulation (e.g., \cite{Martinez2016,Arute2020,Nguyen2022}), we find that the finer local conservation law provided by the gauge symmetry yields a significantly stronger improvement (see Fig.~\ref{fig:postselection_and_calibration}). 
A potential reason is as follows. Two charge violation errors at separate sites, say $i_1$ and $i_2$, can compensate each other to conserve the global charge. In contrast, they will be flagged by non-conservation of the Gauss' law generators $G_{i_1}$ and $G_{i_2}$.

\subsection{Noise model calculations}
\label{sec:noise_calculations}

We use typical calibration data to construct a noise model adapted to the family of chips used for our experiments. This model accounts for Pauli gate errors, thermal noise, coherent two-qubit gate errors, and readout errors. The Pauli noise channels on single- and two-qubit gates are described by mapping the density matrix $\rho$ as
\begin{equation}
    \rho\rightarrow(1-p)\rho+\frac{p}{(4^m - 1)}\sum_{i=1}^{4^m-1}P_i\rho P_i\,,
    \label{equ:depolarizing_error_channel}
\end{equation}
where $p$ is the probability of an error occurring for that individual gate and the $P_i$ are the $4^m-1$ Pauli gates (excluding the identity),  with $m\in\{1,2\}$ the number of qubits partaking in the faulty gate. 

Thermal noise is characterized by the cooling and dephasing operators $\sqrt{1/T_1}\sigma^+$ and $\sqrt{2/T_\upvarphi}n$, where $T_1$ ($T_\upvarphi$) are the qubits' relaxation (dephasing) times and $n=\sigma^-\sigma^+$, resulting in the Lindblad superoperators
\begin{align}
    L_1(\rho)&=\frac{1}{T_1}(\sigma^+\rho \sigma^- - \{n,\rho\}/2)\,,\\\
    L_\upvarphi(\rho)&=\frac{2}{T_\upvarphi}(n\rho n-\{n^2,\rho\}/2)\,,
    \label{equ:thermal_noise_lindbladians}
\end{align}
which are applied after every gate layer taking its execution time into account.

A major source of coherent errors is embodied by extending the idealized two-qubit $\sqrt{\imunit\mathrm{SWAP}}^\dagger$ gates to the excitation-number-conserving generalization
\begin{align}
    &\mathrm{PhasedFSim}(\theta,\zeta,\chi,\gamma,\phi)\nonumber\\
    &=\begin{bmatrix} 1 & 0 & 0 & 0 \\ 0 & e^{-\imunit (\gamma + \zeta)} \cos(\theta) & -\imunit e^{-\imunit (\gamma - \chi)} \sin(\theta) & 0 \\ 0 & -\imunit e^{-\imunit (\gamma + i \chi)} \sin(\theta) & e^{-\imunit (\gamma -\zeta)} \cos(\theta) & 0 \\ 0 & 0 & 0 & e^{-\imunit (2\gamma + \phi)}\end{bmatrix}\,.
    \label{equ:phased_fsim_gate}
\end{align}
In the error-free case, the ideal two-qubit gate is recovered as $\sqrt{\imunit\mathrm{SWAP}}^\dagger=\mathrm{PhasedFSim}(\pi/4,0,0,0,0)$. In our experiments, $\zeta,\chi,\gamma$ are considered negligible after Floquet calibration and correction via single-qubit $z$-rotations, such that the main error derives from the parasitic phases $\phi$ which only affect the subspace of neighbouring excitations.
This renders them of particular relevance for our experiments, which already in the initialisation step host extensive levels of excitations between more than quarter-filling in the computational basis for the case of Fig.~\ref{fig:Overview}a(i) and half-filling for (ii) and (iii).

\begin{figure*}[t!]
    \centering
    \includegraphics[width=\textwidth]{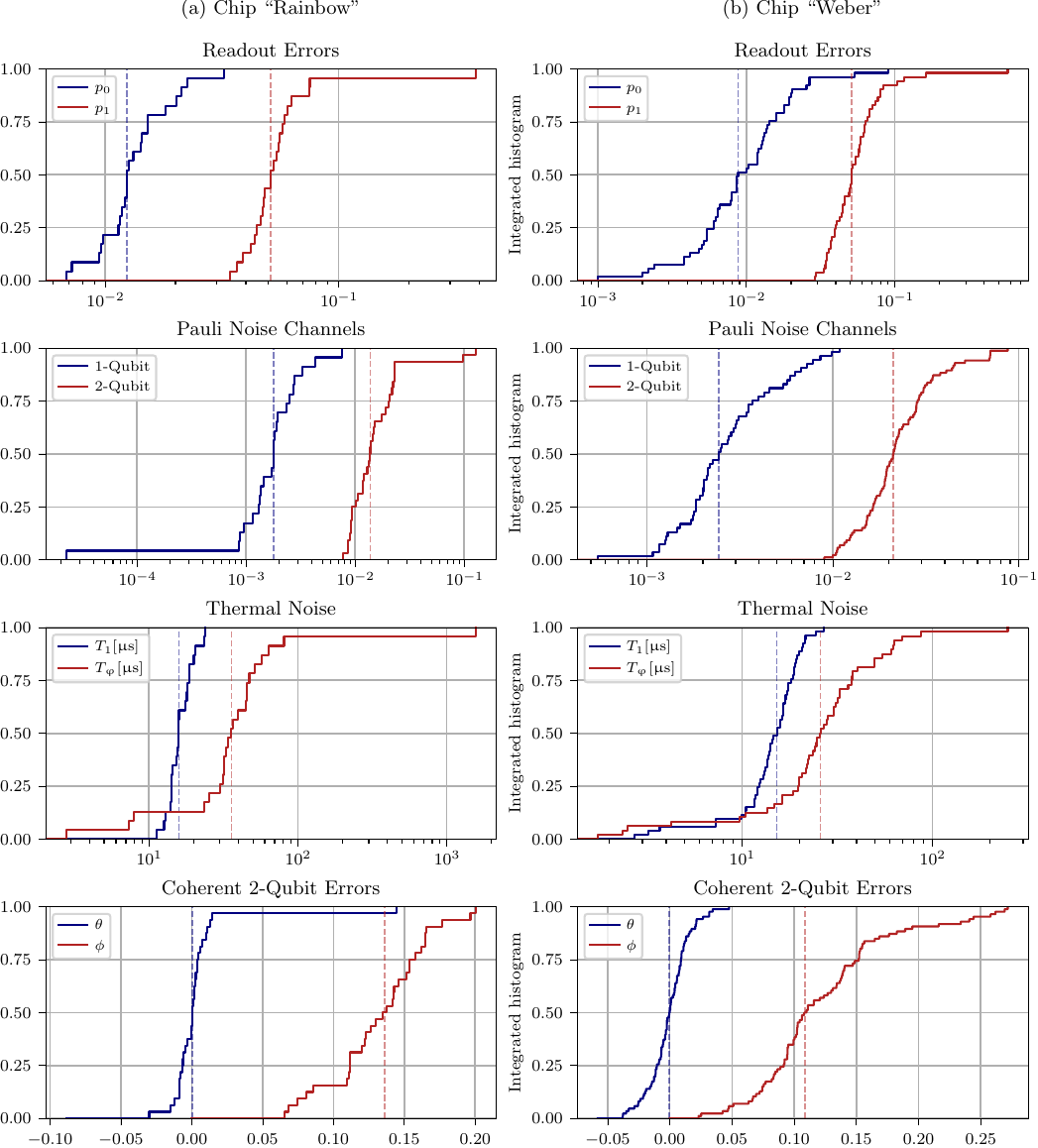}
    \caption{\textbf{Noise Model Parameters.} Integrated histograms showing the values of the parameters constituting the noise model outlined in section~\ref{sec:noise_calculations}, i.e., (from top to bottom) the readout errors $p_0$, $p_1$ used for Eq.~\eqref{equ:readout_error_channel}, the single- and two-qubit errors of the Pauli noise channels in Eq.~\eqref{equ:depolarizing_error_channel}, the relaxation and dephasing times $T_1$, $T_\upvarphi$ in Eq.~\eqref{equ:thermal_noise_lindbladians}, as well as the coherent errors $\theta$, $\phi$ for the native $\sqrt{\imunit\mathrm{SWAP}}^\dagger$-gates introduced in Eq.~\eqref{equ:phased_fsim_gate}. The error data corresponds to (a) the 23-qubit, 32-coupler chip ``Rainbow" and (b) the 53-qubit, 86-coupler chip ``Weber". Dashed lines are median values.}
    \label{fig:noise_histograms}
\end{figure*}

Readout errors are modelled by a generalised amplitude damping channel applied at the end of the circuit, 
\begin{equation}
    \rho \rightarrow M_0 \rho M_0^\dagger + M_1 \rho M_1^\dagger + M_2 \rho M_2^\dagger + M_3 \rho M_3^\dagger\,,
\end{equation}
with
\begin{align}
    &M_0 =\sqrt{p_\mathrm{r}} 
    \begin{bmatrix}
        1 & 0 \\
        0 & \sqrt{1 - \gamma_\mathrm{r}}
    \end{bmatrix},\nonumber\\ 
    &M_1 =\sqrt{p_\mathrm{r}} 
    \begin{bmatrix}
        0 & \sqrt{\gamma_\mathrm{r}} \\
        0 & 0\
    \end{bmatrix},\nonumber\\
    &M_2 =\sqrt{1-p_\mathrm{r}}
     \begin{bmatrix}
        \sqrt{1-\gamma_\mathrm{r}} & 0 \\
        0 & 1
    \end{bmatrix},\nonumber\\ 
    &M_3=\sqrt{1-p_\mathrm{r}} 
    \begin{bmatrix}
        0 & 0 \\
        \sqrt{\gamma_\mathrm{r}} & 0
    \end{bmatrix}\,.
    \label{equ:readout_error_channel}
\end{align}
Here, $p_\mathrm{r}$ and $\gamma_\mathrm{r}$ are determined by the probabilities $p_{0}$ ($p_{1}$) of misreading $\ket{0}$ as $\ket{1}$ ($\ket{1}$ as $\ket{0}$) via $p_\mathrm{r}=\frac{p_{1}}{p_{0}+p_{1}}$ and $\gamma_\mathrm{r}=\frac{p_{1}}{p_\mathrm{r}}=p_{0}+p_{1}$.
The values used for the parameters of the noise model are individual for each qubit/coupler and are taken from actual calibration data representing median performance for this hardware which can be found in Fig.~\ref{fig:noise_histograms}. On a given day, variations with respect to this median may be observed which may also result, e.g., in slightly lower error rates for an experimental session.

With the above noise model, we simulate many trajectories $\ket{\psi_i}$ of the circuits. In order to address inherent inhomogeneities of the used chips, we draw qubits and couplers randomly out of the factorially many possible assignments to a quantum circuit. Each such assignment is used once for all circuit depths.
To qualitatively incorporate the effect of the procedure used in the experiments for the choice of qubits/couplers and assignment averaging, we restrict the set of possible qubits to the better half in the noise model, but in all cases to at least the problem size. In accordance with the one-dimensional layout of our experiments, we also keep the same number of two-qubit gates.

Accounting for the symmetry verification performed in the experiment (see section~\ref{sec:postselection}), we project onto the gauge-invariance and charge conserving symmetry subspace $\mathrm{S}$ as $\Pi_\mathrm{S}\ket{\psi_i}$. From $N$ noise trajectories, the expectation value  of an observable $A$ can be obtained as
\begin{equation}
    \bar{A}=\frac{\sum_{i=1}^N\bra{\psi_i}\Pi_\mathrm{S}\,A\,\Pi_\mathrm{S}\ket{\psi_i}}{\sum_{i=1}^N\bra{\psi_i}\Pi_\mathrm{S}\ket{\psi_i}}\,.	
\end{equation}
This equation may be seen to result from constructing the (frequentist) density matrix based on the trajectories as
\begin{equation}
    \rho=\frac{1}{N}\sum_{i=1}^N\ket{\psi_i}\bra{\psi_i}\,,
\end{equation}
    projecting it into the symmetry preserving subspace
\begin{equation}
    \rho^\prime=\frac{\Pi_\mathrm{S}\rho\Pi_\mathrm{S}}{\mathrm{Tr}(\Pi_\mathrm{S}\rho)}\,,
\end{equation}
and calculating the expectation value $\bar{A}=\mathrm{Tr}(\rho^\prime A)$.
One may compare this to the sampling of measurement results in the experiment, obtaining the expected expression:
\begin{align}
    \bar{A}=\mathrm{Tr}(\rho_\mathrm{SAMP}^\prime A)&=\frac{\sum_{b\in\mathrm{S}}h_b\bra{b}A\ket{b}}{\sum_{b\in\mathrm{S}}h_b}\,,
\end{align}
with
\begin{equation}
    \rho_\mathrm{SAMP}=\sum_{k\in\mathcal{B}}\frac{h_k}{M}\ket{k}\bra{k}\,,
\end{equation}
where $M$ denotes the total number of samples taken and $h_k$ the number of times the measurement returned the result $\ket{k}$ out of the orthonormal measurement basis $\mathcal{B}$.

\begin{figure*}[t!]
    \centering
    \includegraphics[width=\textwidth]{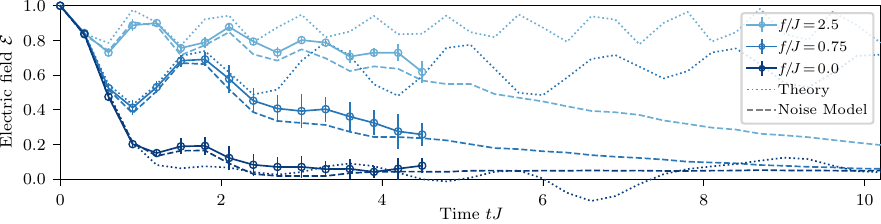}
    \caption{\textbf{Comparison of dynamics of a $\mathbb{Z}_2$ gauge theory in experiment, theory, and noisy simulations towards longer simulation times.} Same as Fig.~\ref{fig:z2_dynamics} of the main text, but extending to longer time scales, including numerical simulations using the comprehensive noise model described in section~\ref{sec:noise_calculations} (dashed lines). Quantitative agreement is found between the dynamics in experimental data (symbols connected by solid lines) and those of exact Trotterized numerical calculations for the noise-free theory (dotted lines) up to the final simulation times displayed in  Fig.~\ref{fig:z2_dynamics} with $tJ\lesssim3$. 
    Oscillations in the experimental dynamics are damped beyond this, but different degrees of decay of the electric field $\mathcal{E}$ as a consequence of the parameter $f$ are discernible on much longer time scales. 
    This effect can both be seen in the experimental data as well as in the well-matching noise model calculations.
    Statistical error bars denote one standard deviation over 10 different circuit assignments around their mean. Data for 8 matter sites plus 8 gauge fields (16 qubits) with periodic boundary conditions, $\mu=0.35J$.}
    \label{fig:z2_dynamics_noisy}
\end{figure*}

\begin{figure*}[t!]
    \centering
    \includegraphics[width=\textwidth]{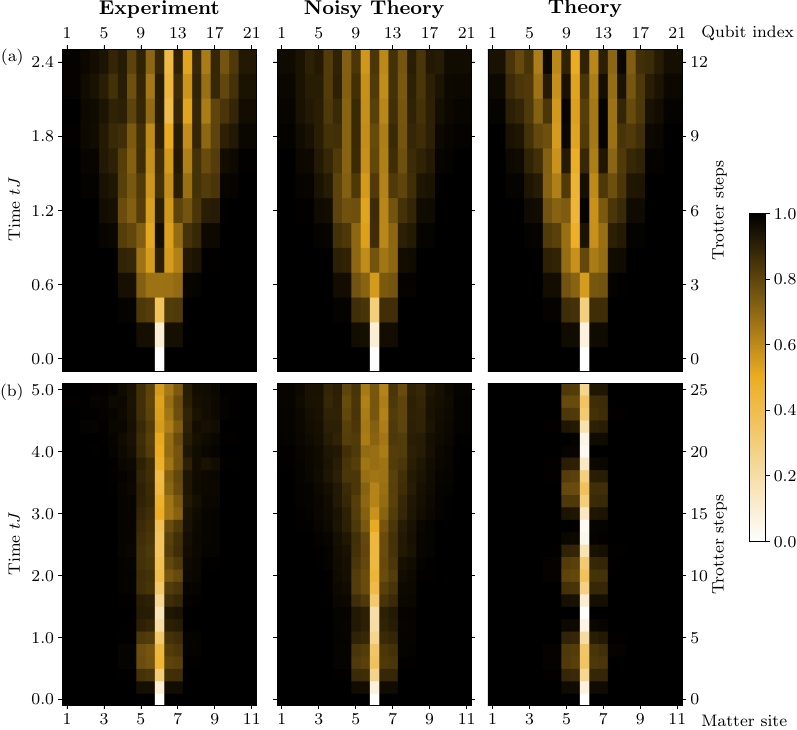}
    \caption{\textbf{Defect dynamics under Noise.} Same as Fig.~\ref{fig:Overview}d,e, but also showing numerical simulations with the noise model given in section~\ref{sec:noise_calculations} (center). The noisy simulations suggest that hardware errors can contribute to a loss of contrast towards the outer edges at late times, as observed in (a) for the case of small $f/J=0.2$ (left) as compared to the noise-free exact Trotterized numerical simulations (right). The additional asymmetry in experimental data can be attributed to systematic errors in 2 of the 5 qubit assignments that were chosen in an automatized way. Over the time scales plotted in most of the figures ($tJ\lesssim 2.5$), the noise model also agrees well with the observations for the case of larger $f/J=2.0$ (b), such as the loss of contrast in experimental data (left) as compared to Trotterized numerical simulation (right). At later times, the noise model even overestimates the impact of errors, indications of which can also be seen in Fig.~\ref{fig:z2_dynamics_noisy}.}
    \label{fig:one_de_noisy}
\end{figure*}

\begin{figure}
    \centering
    \includegraphics[width=\columnwidth]{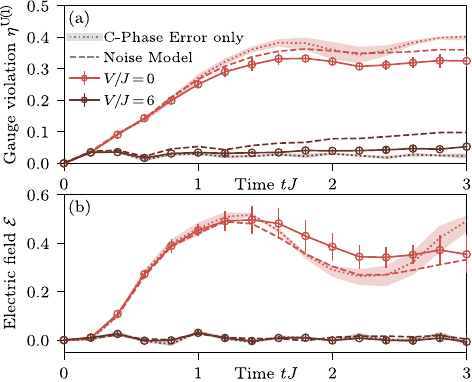}
    \caption{\textbf{Comparison of noise model calculations for freezing of dynamics by gauge-symmetry protection.} Same as Fig.~\ref{fig:u1_protection} of the main text, but also showing exact numerical simulations that only take into account the coherent parasitic C-phase introduced in section~\ref{sec:hardware}, with the hardware-typical value $\phi\simeq0.138\pm0.015$ (dotted lines and shaded region).
    These already explain well leading deviations displayed by the experiment (symbols connected by solid line) compared to theory, yet not as closely as the comprehensive noise model (dashed lines) outlined in section~\ref{sec:noise_calculations}.
    In all cases, postselection on the conserved total matter charge as well as the $\mathbb{Z}_2$ Gauss’s law has been performed as described in section~\ref{sec:postselection}.
    Statistical error bars denote one standard deviation over 10 different circuit assignments around their mean. Data for 6 matter plus 6 gauge fields (12 qubits) with periodic boundary conditions, $f=\mu=2.5J$.}
    \label{fig:u1_protection_noise_comparison}
\end{figure}

As can be seen in Fig.~\ref{fig:z2_dynamics_noisy}, the noise model calculations show good agreement with experimental data. 
In this figure, we have included further data points beyond the final simulation time shown in the main text Fig.~\ref{fig:z2_dynamics}. Although decoherence effects become sizeable at such long times, the noise model captures well the experimentally measured behavior.
Even at long times, the magnitude of the electric field $\mathcal{E}$ is markedly different for different $f$.
The noise model can also partially explain a discrepancy that can be observed between experimental data and theory in Fig.~\ref{fig:Overview}d, namely a loss of contrast towards the edges. As the noise model calculations in Fig.~\ref{fig:one_de_noisy}a (center) display, the hardware errors contribute to this effect. The remaining asymmetry can be explained from systematic errors on the hardware observed in two of the five circuit assignments used. We anticipate that they disappear for circuit averaging over sufficiently many assignments.
Significant effects of decoherence are observed in the longest evolution times reported, i.e., in Fig.~\ref{fig:Overview}e. The corresponding noise model calculations in Fig.~\ref{fig:one_de_noisy}b underline that coherent evolution is maintained approximately up to $tJ\lesssim2.5$. Beyond this time scale, the noise model overestimates experimental errors, which may, i.a., be related to lower-than-median device errors on the chip in the experimental session.
Finally, Fig.~\ref{fig:u1_protection_noise_comparison} shows that a much simpler noise model, i.e., taking into account only a parasitic C-Phase with $\phi\simeq0.138\pm0.015$ as typical for this hardware \cite{Arute2020}, can already explain a substantial part of the deviations of the experimental data from the noise-free theory calculations in Fig.~\ref{fig:u1_protection}, especially at early times. Being a coherent error, however, it misses in particular the damping of oscillations towards later times.

\subsection{Gauge-symmetry protection}

Various methods have been developed in the past to restrain a quantum system to obey a desired symmetry, e.g., by engineered dephasing \cite{Stannigel2014}, by pseudo-random gauge transformations \cite{Lamm2020,Nguyen2022}, by dynamical decoupling \cite{Kasper2023}, or by adding a suitable energy penalty to the Hamiltonian \cite{Zohar2011,Banerjee2012,Zohar2013,Dutta2017,PRL,Yang2020,PRXQuantum}.  
A particularly experimentally friendly approach is given by the concept of \textit{linear gauge protection} \cite{PRXQuantum}, which entails adding to the Hamiltonian a term that is a linear weighted sum in the gauge-symmetry generators and that has a large strength $V$, 
\begin{align}
    \label{eq:VHG}
    H_G=V\sum_ic_iG_i^{\mathrm{U}(1)}\,.
\end{align}
When the coefficients $c_i$ are chosen appropriately, one can effectively turn the gauge superselection sectors (defined by the eigenvalues of the $G_i^{\mathrm{U}(1)}$) into quantum Zeno subspaces that remain uncoupled up to a timescale at least linear in $V$ \cite{Facchi2002,Disorder-Free_Localization_1}.

Even more, when working in a single target gauge superselection sector, defined through the conserved charges $\{g_i\}=\{g_i^\text{tar}\}$, it suffices to only isolate this sector from the rest of the Hilbert space. When the sequence $c_i$ is \textit{compliant}, i.e., it satisfies
\begin{align}\label{eq:compliance}
    \sum_ic_i\big(g_i-g_i^\text{tar}\big)=0\iff g_i=g_i^\text{tar},\,\forall i\,,
\end{align}
the effective protection of the target gauge sector is guaranteed up to timescales exponential in the protection strength $V$ \cite{Abanin2017,PRXQuantum}. 

\begin{figure}
    \centering
    \includegraphics[width=\columnwidth]{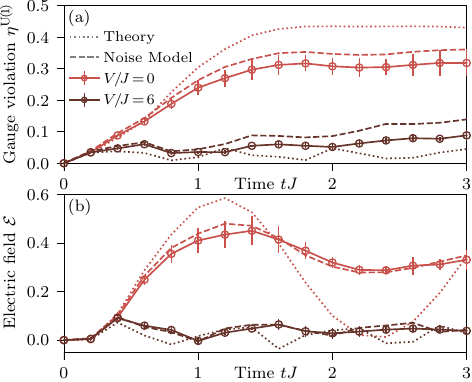}
    \caption{\textbf{Gauge protection and freezing of dynamics induced by modified local conservation law with simple protection sequence.} The scenario is as in main-text Fig.~\ref{fig:u1_protection}, but for 8 matter plus 8 gauge fields (16 qubits) with periodic boundary conditions and the simple noncompliant sequence $c_i=\pm1$, which nevertheless leads to satisfactory protection of $\mathrm{U}(1)$ gauge symmetry over the evolution time.
    Statistical error bars denote one standard deviation over 10 different circuit assignments around their mean.}
    \label{fig:u1_protection_extended}
\end{figure}

The relation~\eqref{eq:compliance} ensures that no combination of processes that violate the conservation of $g_i^\text{tar}$ at different matter sites can be resonant.  
In principle, the sequence $c_i$ will have to grow exponentially with system size in order to satisfy this relation. Interestingly though, depending on the dominant terms that couple the gauge superselection sectors, much simpler \textit{noncompliant} sequences can already suffice to stabilize gauge invariance up to all relevant timescales \cite{PRXQuantum,Vandamme2023}. 
Fig.~\ref{fig:u1_protection_extended} shows the results analogous to Fig.~\ref{fig:u1_protection} of the main text for 16 qubits (8 matter sites, 8 gauge-field links) and the very simple sequence $c_i=(-1)^{i+1}$, which already suffices to produce satisfactory results. 
In the main text Fig.~\ref{fig:u1_protection}, we use the sequence $c_i\in\{-115,116,-118,122,-130,146\}/146$, which is compliant for a system of $N=6$ matter sites. In larger systems, this sequence ensures that no combination of processes that violates conservation of $G_i^{\mathrm{U}(1)}$ is resonant within a distance of $N=6$ matter sites. 

A main feature of Eq.~\eqref{eq:VHG} is that it involves only terms linear in the gauge-symmetry generators $G_i^{\mathrm{U}(1)}$, enabling its implementation using only single-qubit terms. These incur no relevant experimental overhead in our implementation as they can be absorbed into already present terms. 

A subtlety of the application of $H_G$ comes from Trotterization \cite{PRXQuantum}. In principle, one would want to make the associated scale $V$ as large as possible in order to improve the level of $\mathrm{U}(1)$ gauge symmetry. However, in a Trotterized sequence the protection terms $\exp(\imunit\Delta tVc_iG_i^{\mathrm{U}(1)})$ are periodic under $\Delta t V c_i$. Consequently, there is an optimal strength $V^*$ beyond which protection no longer improves (conveniently for implementations, no fine tuning is required since a broad range of $V$ around $V^*$ yields comparable performance  \cite{PRXQuantum}). 
For our parameters, $V^*$ is close to $V/J=6$. Importantly, as seen in Fig.~\ref{fig:u1_protection} we find that this moderate value already suffices to efficiently induce an approximate $\mathrm{U}(1)$ gauge symmetry.

\end{document}